# Self-limiting growth of two-dimensional palladium between graphene oxide layers

Yang Su[1,2], Eric Prestat[3], Chengyi Hu[1,2], Vinod Kumar Puthiyapura[1,2], Mehdi Neek-Amal*[,4,5], Hui Xiao[1,2], Kun Huang[1,2], Vasyl G. Kravets[6], Sarah J. Haigh[3], Christopher Hardacre[2], Francois M. Peeters[4], Rahul R. Nair*[,1,2]

[1]National Graphene Institute, University of Manchester, Manchester, M13 9PL, UK.

[2]School of Chemical Engineering and Analytical Science, University of Manchester, Manchester, M13 9PL, UK.

[3]School of Materials, University of Manchester, Manchester M13 9PL, UK.

[4]Department of Physics, University of Antwerpen, Groenenborgerlaan 171, B-2020 Antwerpen, Belgium.

[5]Department of Physics, Shahid Rajaee Teacher Training University, Tehran, Iran.

[6]School of Physics and Astronomy, University of Manchester, Manchester M13 9PL, UK.

**ABSTRACT : The ability of different materials to display self-limiting growth has recently attracted enormous attention due to the importance of nanoscale materials in applications for catalysis, energy conversion, (opto)electronics, etc. Here, we show that electrochemical deposition of palladium (Pd) between graphene oxide (GO) sheets result in a self-limiting growth of 5 nm thin Pd nanosheets. The self-limiting growth is found to be a consequence of strong interaction of Pd with the confining GO sheets which results in bulk growth of Pd being energetically unfavourable for larger thicknesses. Furthermore, we have successfully carried out liquid exfoliation of the resulting Pd-GO laminates to isolate Pd nanosheets and demonstrated their high efficiency in continuous flow catalysis and electrocatalysis.**

GO laminate with nanometer and sub-nanometer size interlayer channels is continuously receiving tremendous attention due to its unique molecular permeation properties[1-5]. In particular, tunable ion, molecular sieving and fast mass transport[4-8] make it a promising candidate for next-generation membrane for advanced separation technologies[1,9,10]. Even though the molecular transport through nanometer wide channels in GO laminates has been extensively studied[2-8], the use of nanochannels for confinement-controlled chemistry to synthesize novel nanomaterials is not explored. It has been previously demonstrated that nano-confinement can hugely influence the chemical reactions at nanoscale[11-14]. There have been several attempts in using GO as a template to grow metal oxide nanosheets by calcining the corresponding salt adsorbed laminates[15-18]. However, in such a process, the uncontrollable expansion of the interlayer space during the calcination process nullifies the confinement effect, and leads to the formation of nanosheets with different morphologies and thickness. Herein, we show that electrochemical deposition of Pd between GO sheets result in a self-limiting growth of 5 nm thin Pd nanosheets. The thickness of the nanosheets deposited between the GO layers is found to be independent of growth time, and the growth self-terminates after reaching the full coverage to the contacting GO layers. Considering that the self-limiting growth of materials has only been known for surface growth of an atomic or molecular layer on specific substrates[19-21], the reported self-limiting growth between GO layers will open new prospects in self-limiting crystal growth.

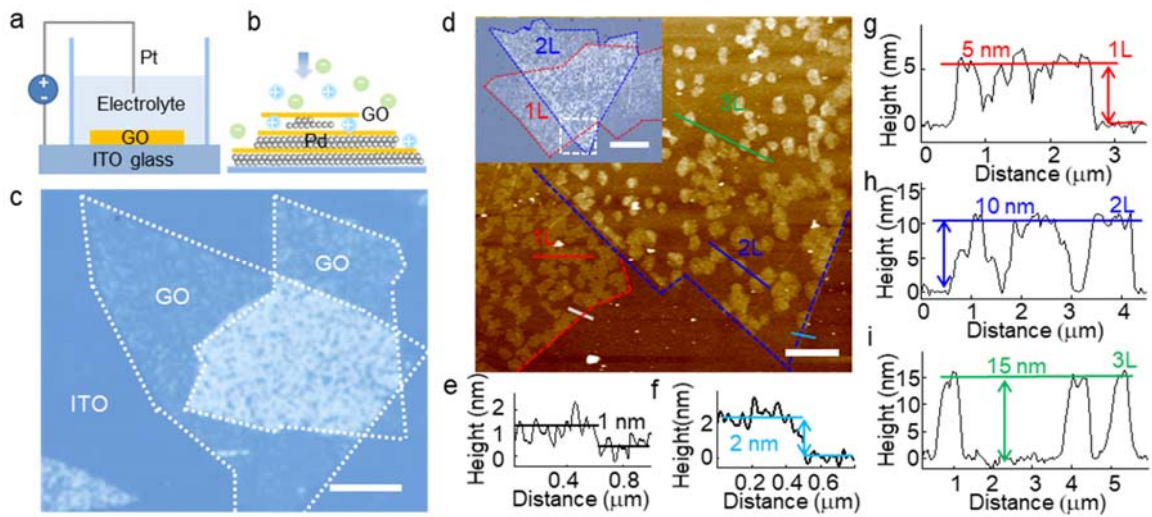

**Figure 1. Electrochemical growth of Pd between GO layers. a**, Schematic of the ECD experimental setup. **b**, Schematic of ECD of Pd between GO layers. **c**, Optical image of two GO sheets on an ITO surface after 30 seconds ECD. The dotted lines profile the two GO sheets. Scale bar, 5 μm. **d**, AFM image obtained from the overlapped single layer and double layer GO sheets after the ECD. The red and blue dotted lines profile the edges of the single and bilayer GO sheets respectively. Scale bar, 2 μm. Inset: optical micrograph of the corresponding GO sheets. Red and blue dotted lines profile the single layer and bilayer GO sheets respectively. The white rectangle shows the area from which the AFM image was obtained. Scale bar, 20 μm. **e-f**, AFM height profiles of the GO sheets along the black and light blue lines in Figure 1d. **g-i**, AFM thickness profiles (colour coded) along the red, blue and green lines in Figure 1d showing the thickness of the Pd nanosheets deposited between GO layers scales with the number of GO layers.

Figure 1a shows a schematic of the experimental set-up we used for the electrochemical deposition (ECD) of Pd between the GO sheets. We used a two-electrode ECD configuration, consisting of a platinum anode, GO coated indium tin oxide (ITO) glass cathode and an aqueous Pd(NO$_3$)$_2$ solution (1 M) as the electrolyte (Methods). GO flakes were deposited on the ITO glass by a drop casting technique. ECD was carried out using a Keithley 2400 SourceMeter at a current density of 0.1-0.5 mA/cm$^2$. During the ECD process, we kept the applied voltage below the reduction potential of GO sheets (< 1 V) to avoid possible electrochemical reduction of GO[22,23].

Figure 1c is an optical microscopy image of two overlapped GO flakes on ITO glass after the ECD for 30 seconds. It shows three different levels of contrast: featureless ITO glass, GO flake with bright spots, and the two overlapped GO flakes with the highest contrast. Even though ITO glass is a good electrical conductor, the absence of any features on the ITO substrate and the presence of bright spots on the electrically insulating GO suggest preferential Pd nucleation in areas covered by the GO layer. This is further confirmed by scanning electron microscope (SEM) (Figure S1 and supplementary section 1). The higher contrast from overlapped GO flakes also indicates that Pd growth is more preferred on the multilayer GO compared to a single or fewer layer material, suggesting the crystal growth is occurring in the interlayer space between the GO sheets. To probe this further, we have performed atomic force microscope (AFM) measurements on the GO sheets after the ECD of Pd. Figure 1d shows an AFM image containing a single layer, a bilayer, and an overlapped single and bilayer region (three layers) after the ECD of Pd. Thickness profiles across the edge of the single and bilayer GO are shown in Figure 1e and f, respectively. The AFM images clearly reveal the 2D flaky nature of the deposited Pd, and more interestingly the thickness profiles accross the metal nanosheets taken from different thick GO regions (Figure 1g-i) show a scaling behavior. That is, Pd nanosheets present on the monolayer GO region are ~ 5 nm thick (Figure 1g), while, on bilayer they are 10 nm (Figure 1h), and on the three-layered GO sheets they are 15 nm (Figure 1i). This scaling behaviour suggests that the Pd nanosheets are deposited not only at the interface between GO and the ITO cathode, but also at the interlayer spaces between GO layers. These AFM measurements also indicate that the Pd crystals starts to nucleate at the ITO - GO interface and the formed Pd acts as an extended electrode for the growth of additional Pd layers between the GO sheets laying above (Figure 1b).

Such growth events repeat until the Pd growth reaches the top GO surface after which the growth then follows normal particulate growth as in the case of ECD of Pd on any conducting substrates such as bare ITO or few-layer graphene or reduced GO (rGO) (Figure S1).

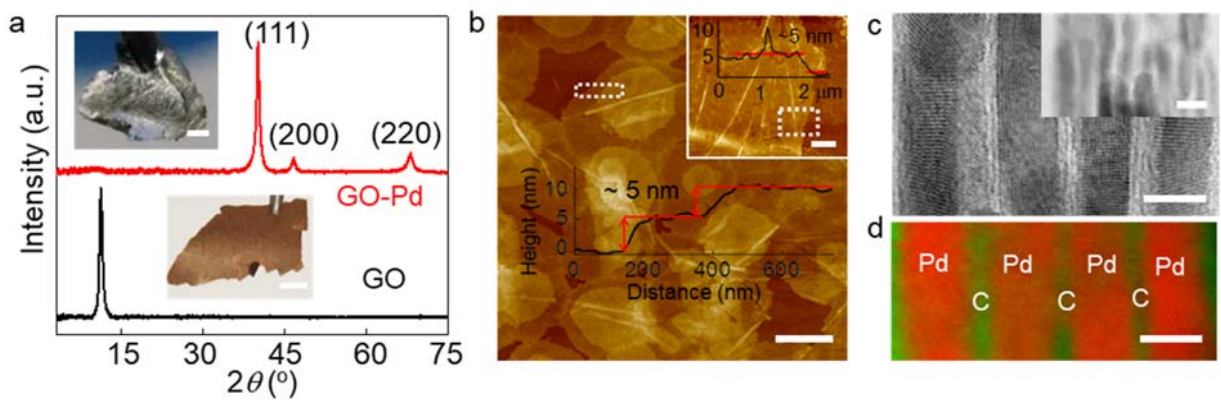

**Figure 2. Electrochemical growth of Pd inside the GO laminate. a**, XRD for GO laminate before (black) and after (red) the ECD process for 24 hours. Inset: (lower) optical micrograph of GO laminates before and (upper) after the ECD. Scale bars: 2 mm. **b**, AFM image obtained from a GO laminate after the ECD for 1 hour exfoliated on a silicon substrate. Scale bar, 500 nm. Lower inset: the height profile along the dotted rectangle. Upper inset: AFM image showing the continuous Pd layer close to the cathode and its thickness profile along the white rectangle. Scale bar, 1 μm. **c**, Cross-sectional HRTEM image of a GO laminate after 24 hours of ECD. Scale bar, 5 nm. Inset: the low magnification TEM of the cross section of a GO laminate after 24 hours of ECD. **d**, STEM-EELS elemental mapping of the cross-section in Figure 2c. Scale bar, 5 nm.

To further study the preferential growth of Pd between the interlayer spaces of the GO sheets and to demonstrate the scalability of this method, we have performed ECD on thick GO laminates (0.2 - 3 μm thick) (methods). During the ECD, the growth time was carefully controlled to avoid any possible surface growth of Pd outside the GO laminate. As shown in Figure 2a, after the ECD, the GO laminate changed colour from a semi-transparent brown to a shiny metallic grey. The X-ray diffraction (XRD) spectra show that the peak corresponding to the (001) reflection of the GO laminate nearly disappears after the ECD, while the new peaks corresponding to metallic Pd appear, confirming the successful growth of Pd inside the GO laminate. To characterize the deposited Pd inside the laminate, we have cleaved the Pd deposited GO laminate using scotch tape to expose the inner surfaces of the laminate, and performed AFM measurements. We have found two types of Pd nanosheets in the GO laminate. (i) scattered Pd nanosheet islands with a thickness of ~ 5 nm (Figure 2b) found close to the top surface of the laminate and (ii) continuous layers of Pd with a step size of 5 nm, which were found at the freshly cleaved surface close to the cathode, towards the bottom of the GO laminate (Figure 2b inset). It is noteworthy that the Pd islands were not overlapped laterally, each island appears to grow until it reaches the boundary of next island with a self-limiting growth behavior. Such growth also could explain the continuous 5 nm Pd layers found close to the cathode. Since the Pd growth initiates at the cathode and continues to grow towards the top side of the laminate, different Pd islands in the GO interlayers close to the cathode quickly merge to form a continuous Pd layer. To further confirm this self-limiting growth of Pd between GO sheets we have performed cross-sectional transmission electron microscopy (TEM) on Pd deposited GO laminates.

For the TEM, ECD was performed for a longer time (24 hours on a 2 μm thick sample) to make sure that Pd growth is complete accross the full interlayer spaces of the GO laminate. Figure 2c shows the cross-sectional TEM image of the Pd deposited GO laminate revealing a brick wall or nacre structure. Electron energy loss spectroscopy (EELS) elemental mapping of the cross-section further provides direct evidence for the alternating layers of Pd and carbon in the laminate (Figure 2d). Figure 2c show the high-resolution TEM (HRTEM) images of the cross-section further revealing the polycrystalline nature of the Pd layers, consistent with the XRD measurement. Each Pd layer is found to be ~ 5 nm, which is consistent with our AFM measurements and supports the self-limiting growth behaviour. The number of GO layers that separate the adjacent Pd layers varies from 1 to 6 as measured for different regions of the sample. The varying thickness of the GO spacers could be explained by the weak electric field screening of thin GO layers (≈ < 6 layers), which as a result bypasses the growth of Pd on these layers.

The formation of a Pd between the GO layers offers the prospect of isolating the Pd layers in a water-based solution. To achieve this, we have followed a liquid exfoliation technique (Figure S2)[24]. We noticed that the chemical reduction of the GO-Pd laminate using 50% hydrazine solution at 80 $^0$C for 30 min, improves the exfoliation process due to the reduction of GO causing bubble formation at the GO-Pd interface. Figure 3a shows a photograph of the dispersion obtained after exfoliation of the Pd-rGO laminate in a water and isopropanol mixture. AFM measurements of the isolated Pd nanosheets show that ~ 20% of the nanosheets are ~ 5 nm thick and the rest are multiples of 5 nm (Figure S3). The HRTEM image along with the corresponding selected area electron diffraction (SAED) pattern obtained from the exfoliated material confirm the crystalline nature of the Pd nanosheets (Figure 3b-c) and prove the absence of any oxidation of the Pd.

To explore the potential applications of the synthesized Pd nanosheets, we have performed two sets of experiments. First, we studied the catalytic properties of Pd nanosheets by re-assembling them into a membrane for continuous flow catalysis to convert 4-nitrophenol to 4-aminophenol using $NaBH_4$ as reductant, which is a model reaction for examining the catalytic properties of metal catalysts[25]. It has been previously suggested that a membrane-based continuous flow catalysis approach could avoid the time- and energy-consuming recycling process which is required when catalyst nanoparticles are used also and provide high catalytic conversion efficiency[26,27]. Pd membranes were prepared by reassembling the exfoliated Pd nanosheets/rGO on a porous poly(ether sulfones) support by vacuum filtration (Figure 3d). Note that, we also added 10-20 wt% GO to the Pd dispersion to form a laminate membrane which allows permeation through the tortuous Pd interlayer nanochannels (Figure S2). Without the addition of GO, the flux was high, but no catalytic conversion was observed, suggesting the permeation mainly occurs through the defects piercing through the membrane and not between the layers of Pd needed to activate the catalytic conversion. Also, a pristine GO membrane with similar thickness shows negligible 4-nitrophenol permeance, which is consistent with previous reports[28]. The catalytic membrane conversion experiments were performed using a vacuum filtration set-up[27]. For a 2 μm thick Pd nanosheet membrane, we have obtained stable permeance of ~ 80 L h$^{-1}$ m$^{-2}$ bar$^{-1}$ (Figure 3e inset). The conversion of 4-nitrophenol to 4-aminophenol was studied by using UV-Vis absorption spectra since 4-nitrophenol has its characteristic absorption peak at ~400 nm and 4-aminophenol shows the characteristic peak at ~300 nm[25]. Figure 3e shows the absorption spectra for both the permeate and the feed solution, clearly demonstrating the complete conversion of 4-nitrophenol to 4-aminophenol by the Pd nanosheet membrane. Secondly, we studied the potential of Pd nanosheets to act as a catalyst for the methanol oxidation reaction and have demonstrated its superior performance over commercial Pd black (supplementary section 2 and Figure S4).

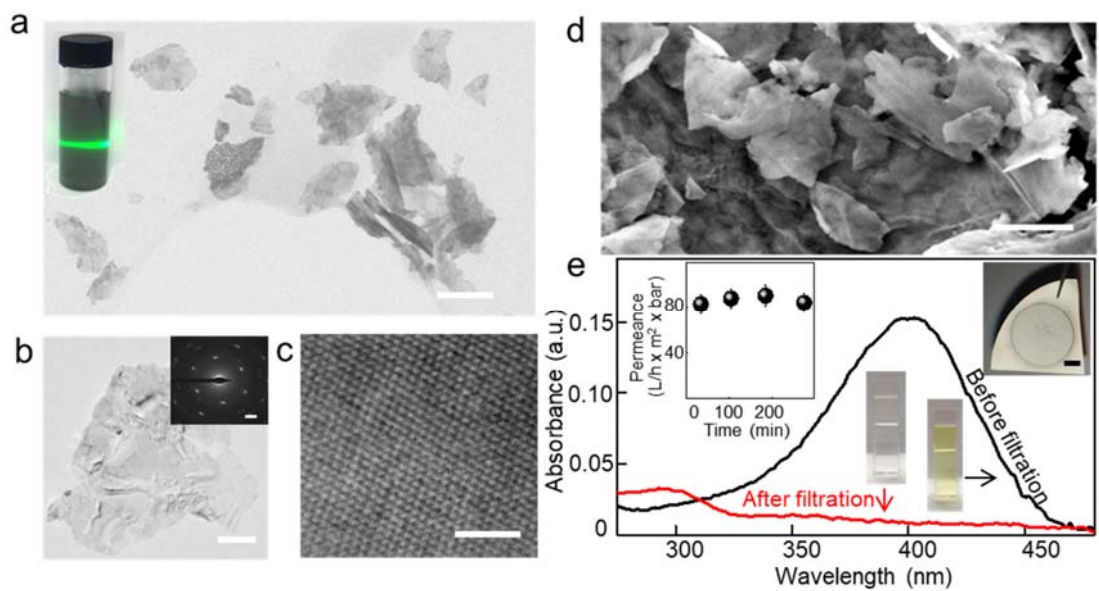

**Figure 3. Exfoliation of Pd nanosheets and their catalytic applications. a**, Low magnification TEM image of Pd nanosheets exfoliated from the Pd deposited GO laminate. Scale bar, 1 μm. Inset: optical photograph of the colloidal solution of exfoliated Pd/rGO. A green laser beam was directed through the dispersions to show the Tyndall effect, which confirms their colloidal nature. **b**, Representative TEM image of the exfoliated Pd nanosheet. Scale bar, 100 nm. Inset: selected area electron diffraction pattern from Figure 3**b**. Scale bar, 2 1/nm. **c**, HRTEM image of the Pd nanosheet. Scale bar, 2 nm. **d**, SEM image of Pd membrane used for continuous flow catalysis. Scale bar, 400 nm. **e**, UV-vis optical absorption spectrum of 4-nitrophenol before (black) and after (red) filtration through the Pd membrane. Inset (top left): permeance of 4-nitrophenol through Pd membrane as a function of time. Inset (top right): Optical photograph of a Pd membrane. Scale bar: 5 mm. Inset (bottom): photographs of the solution of 4-nitrophenol before and after the filtration through the Pd membrane.

The observed self-limiting growth of 2D Pd nanosheets between GO layers is intriguing and is fundamentally interesting as an unusual crystal growth phenomenon. To explain the self-limiting growth, we adapted the classical nucleation theory used to describe the growth of spherical nanoparticles[29]. It is known that the total free energy $\Delta G$ of the nucleus dominates the crystal nucleation process and depends on the surface free energy and the bulk free energy. However, for nucleation in a situation of 2D confinement (between GO sheets with ~ 1 nm spacing), the interaction between particle and sheets, as well as the van der Waals adhesion between the two layers needs to be considered. By modeling the nucleus as a truncated spherical segment with a radius $r$ and height $z$ (Figure S5), $\Delta G$ can be expressed as,

$$\Delta G(r, z) = Vu(z) + S\gamma + (S_1 + S_2)\gamma' + E_{vdW}(z) \qquad (1)$$

where $r$ is the radius of the nucleus, $z$ is the height/thickness of the nucleus, $V$ and $S$ are the volume and accessible surface area for the growing nucleus, $u$ is the cohesive energy per unit volume of the nucleus between the GO sheets, $S_1$ and $S_2$ are the nucleus' contact areas with the top and bottom GO layers respectively, $\gamma$ is the surface free energy, $\gamma'$ is the interaction between GO sheets and Pd (adhesion energy), and $E_{vdW}(z)$ is the van der Waals adhesion energy between two GO sheets (supplementary section 3). Previously it has been found that Pd interacts with graphene via a strong chemisorption and $\gamma'$ was estimated as -0.084 J/m² [30]. Typically for GO, an area of 40–60% remains free from functionalization[31,32], hence we assume that $\gamma'$ does not significantly vary from pristine graphene. However, it is noteworthy that the oxygen functional groups at the GO surface has a role of reducing the electrical conductivity of GO laminate otherwise Pd nucleation could have occur at the surface of GO laminate similar to the electrodeposition od Pd on pristine graphene or reduced GO samples (Figure S1). By calculating $\Delta G$ for different values of $r$ and $z$, we found that for $z < \sim 5$ nm, $\Delta G$ becomes $< 0$ and for $z > 5$ nm, $\Delta G$ becomes $> 0$ for all values of $r$. This suggests that Pd growth between the GO sheets is not favorable if their thickness is $> 5$ nm (Figure S5). Thus, we propose that the observed self-limiting growth of Pd is a consequence of the strong interaction of Pd with the GO and the 2D confinement created by the interlayer space. To further validate this proposed mechanism, we have performed ECD of copper (Cu) between GO layers and found that even though Cu has a 2D flaky structure, the thickness was varying from several to tens of nanometers, suggesting the absence of self-limiting growth (see supplementary section 4 and Figure S6). This could be explained by the weaker interaction of Cu with graphene compared to that of Pd.

In conclusion, we have shown that electrochemical growth of Pd between the interlayer spaces of GO leads to a self-limiting growth of 5 nm thin Pd nanosheets. The self-limiting growth we observe is attributed to the interlayer confinement and strong interaction of Pd with the GO layers. Taking into account the highly favourable catalytic properties of Pd, we have demonstrated the potential application of Pd nanosheets for continuous flow catalysis, and electrocatalysis. Our work shows a novel route to the production of 2D nanosheets using 2D laminates as a host and invites for broad explorations.

**Methods**

**Preparation of graphene oxide:** Graphite oxide was prepared by the modified Hummers' method as reported previously[28]. An aqueous solution of GO with a concentration of 1 mg/mL and a pH ~ 7 was prepared by dispersing graphite oxide flakes in distilled water using bath sonication followed by centrifugation to remove the multilayer GO flakes.

**Electrochemical deposition**: ECD experiments were performed using GO deposited ITO substrates (sheet resistance ~200 Ω/square), or 10 nm Al/30 nm Au coated free standing GO laminates as the cathode. Platinum electrode (99.9%, 25 μm thick platinum foil with 5 mm width) was used as the anode. For the fabrication of scattered GO flakes on ITO, diluted GO suspension with a concentration ~ 1 μg/mL was drop casted on the ITO substrates. For thicker GO laminate electrodes, we deposited ~200 nm thick GO laminate on the ITO substrate by drop casting GO solution with a concentration of 0.1 mg/mL. Laminates with higher thickness were prepared by vacuum filtration and 10 nm Al/30 nm Au was thermally evaporated on one side of the laminate and used as an electrode.

Electrolyte reservoir for the ECD was fabricated by attaching an open-end plastic tube onto the GO electrode using an epoxy resin. Before the ECD, the GO sheets or laminates were stabilized in the electrolyte for several minutes to allow the electrolytes to permeate into the GO laminates. The platinum anode was kept at a distance ~ 5 mm away from the GO electrode during the ECD. For the ECD, both the anode and cathode were connected to the Keithley 2400 SourceMeter, and the current density was controlled at 0.1-0.5 mA/cm$^2$. The electrolyte concentration for all the ECD reported here was fixed at 1 M. We have also studied the influence of electrolyte concentration on the Pd deposition and found that 0.5 M and higher electrolyte concentration provides 5 nm Pd nanosheets whereas lower electrolyte concentration leads to Pd nanoparticle formation. The particle formation could be due to the lack of sufficient Pd ions between the GO sheets at lower electrolyte concentration which limits the growth of the Pd nucleation sites.

**Characterization of electrodeposited Pd between GO sheets:** The individual GO flakes on ITO, after the ECD, were rinsed with deionised water and directly used for AFM and SEM studies. In the case of Pd nanosheet incorporated GO laminates (Pd-GO laminate), after ECD, the samples were randomly peeled off by scotch tape to expose the fresh surface of Pd crystals grown inside the GO laminate. The thicknesses of Pd nanosheets were measured using a Veeco Dimension 3100 AFM in the tapping mode and SEM characterisations were conducted in a Zeiss ultra 55 microscope. The XRD measurements in the 2θ range of 3° to 75° (with a step size of 0.02° and recording rate of 0.2 s) were performed using a Bruker D8 diffractometer with Cu Kα radiation (λ = 1.5406 Å). For the TEM cross-sectional sample preparation, the Pd-GO laminates were glued between two pieces of Si wafer and 500 μm thick section were extracted using a wire saw. The section was mechanically polished in to the shape of a wedge down to electron transparency using an Allied multiprep polisher system. Finally, the section was polished using low-energy (0.2-1 keV) $Ar^+$ ion to achieve thinner specimen thicknesses. TEM characterisation was carried out in a FEI Titan ChemiSTEM microscope operating at 200 kV. EELS experiments were performed using a Gatan Imaging Filter (GIF) Quantum. The C and Pd maps were obtained from the C-K edge and Pd-M edge, respectively. For both edges, the pre-edge background was subtracted using a power law model and the remaining edge intensity was integrated over a range of 50 eV after the edge.

**Exfoliation of Pd nanosheets and their catalytic properties:** We used the Pd-GO laminate for the exfoliation of Pd nanosheets. The samples were obtained after a long-time ECD (for thin membranes on ITO, we deposited for 1 hour, and for thick free-standing membranes, a 24 hours deposition was used). The laminate after long-time ECD was peeled off from the ITO substrate by immersing it in water. For the thick GO laminates, the 10 nm Al/30 nm Au is etched and lifted off by 1 M hydrochloric acid. The Pd-GO laminates were then reduced in 50% hydrazine solution at 80 $^0$C for 30 min, repeatedly washed with water and dried in a vacuum oven at 100 $^0$C. The exfoliation was conducted by immersing the dry Pd/rGO in a mixture of isopropanol and water (volume ratio is 1:1). After 48 hours of ultra-sonication, the resultant colloid was directly used for TEM and AFM measurement by drop casting the colloid on a gold TEM grid and silicon wafer respectively. We have also exfoliated the GO-Pd laminates without the hydrazine reduction process, but the exfoliation yields were found to be smaller.

For the continuous flow conversion of the 4-nitrophenol experiment, a 2.5 mL Pd/rGO colloid with a concentration of 1 mg/mL was mixed with 0.5 mL of 1 mg/mL GO suspension followed by filtration through a porous polysulfone polymer support (pore size: 0.22 μm, Millipore). 100 mL of 4-nitrophenol (0.1 mM) solution mixed with 4 mL of 56 mM $NaBH_4$ solution was used as feed solution. The permeance was monitored at different time intervals to make sure that the experiment is under a steady state. UV-visible-near-infrared grating spectrometer with a xenon lamp source (250-2500 nm) was used for monitoring the conversion of 4-nitrophenol. In this measurement, relative transmission spectra were measured as the ratio of sample (feed and permeate) transmission to that of the pure cuvette and then converted into absorbance.

ACKNOWLEDGMENT

This work was supported by the Royal Society, Engineering and Physical Sciences Research Council, UK (EP/S019367/1, EP/P025021/1, EP/K016946/1 and EP/P009050/1), Graphene Flagship, and European Research Council (contract 679689 and EvoluTEM). We thank Dr. Sheng Zheng and Dr. K. S. Vasu at the University of Manchester for assisting with sample preparation and characterisations. The authors acknowledge the use of the facilities at the Henry Royce Institute for Advanced Materials and associated support services. V.K.P, and C.H kindly thanked for the resources and support provided via membership of the UK Catalysis Hub Consortium and funded by EPSRC (Portfolio Grants EP/K014706/2, EP/K014668/1, EP/K014854/1, EP/K014714/1, and EP/I019693/1). F.M.P, and M.N acknowledges the support from Flemish Science Foundation (FWO-Vl).

# Supplementary Information

1.  **Electrochemical deposition of Pd on graphene oxide, few layer graphene and reduced graphene oxide covered indium tin oxide (ITO) glass substrate.**

To confirm the preferential self-limiting electrochemical growth of Pd between graphene oxide (GO) layers, we have carefully examined the Pd growth on a partially GO covered ITO glass substrate by using scanning electron microscopy (SEM). Figure S1a shows a SEM image of a partially covered GO on ITO glass after the electrochemical deposition (ECD) of Pd for 30 seconds. As evident from the image, only a small fraction of Pd is deposited on bare ITO in comparison to the GO covered ITO where a denser and flaky Pd deposition can be seen, confirming the preferential Pd growth between the GO layers. To further understand the influence of graphene surfaces on the morphology of the deposited Pd, we have also performed ECD on reference substrates such as mechanically exfoliated pristine few-layer graphene (FLG) on ITO glass and reduced GO (rGO) flakes (obtained by hydrazine vapour reduction[1]) on ITO glass. Figure S1b and c show SEM images obtained from FLG and rGO respectively, after the ECD of Pd. Both the substrates show particulate growth of Pd in contrast to the two dimensional (2D) flaky Pd between GO sheets confirming the importance of the GO capillaries on the observed self-limiting growth of Pd. It is also noteworthy that due to the high conductivity of graphene and rGO, no Pd deposition was observed between the layers of graphene and rGO. In addition, the importance of the GO capillaries on the self-limiting growth can be further evidenced by the three-dimensional morphology of Pd grown outside the GO layers after an extended ECD of Pd (Figure. S1d).

2.  **Electrocatalytic performance of the Pd sheets - Methanol oxidation reaction (MOR)**

We studied the potential of the synthesised Pd nanosheets as a catalyst for MOR by comparing its performance with respect to the commercial Pd black (Sigma-Aldrich, surface area: 40-60 $m^2/g$). For the preparation of the working electrode used in electro-catalysis experiments, the exfoliated Pd colloid and Pd black ink were drop-cast on the glassy carbon electrode (diameter: 5 mm). The catalyst ink was prepared by sonication of the Pd sample in a mixture of water, isopropanol (volume ratio 1:3) and Nafion solution. Nafion acts as binder to prevent the catalyst from peeling off during the reaction. All the sample loadings were controlled to be ~100 μg/$cm^2$.

The electrochemical measurements were carried out in a three-electrode glass cell connected to a Bio-Logic EC Lab SP-200 potentiostat. A calomel (3M KCl) electrode and a Pt mesh were used as the reference and counter electrode respectively. All potentials are reported with respect to the potential of normal hydrogen electrode (3M KCl calomel = 0.254 V vs NHE at 25 ºC).

We first determined the electrochemical surface area (ESA) of Pd sheets and Pd black by using cyclic voltammetry (CV). This was carried out in $N_2$ saturated 0.5 M $HClO_4$ solution at a scan rate of 50 mV/s in the potential range 0.054 V to 1.414 V. ESA of Pd was calculated using Eq. 1 where $Q_o$ is the charge required for the PdO reduction which was derived from the PdO reduction peak in 0.5 M $HClO_4$ solution, and $Q_{PdO}$ is the charge required for the reduction of a monolayer of PdO (405 μC/cm$^2$)[2,3,4].

$$ESA = \frac{Q_o}{Q_{PdO}} \qquad (1)$$

The catalytic methanol oxidation on Pd black and Pd sheets were compared by CV measurements in 0.1 M methanol + 0.5 M KOH solution. The current was normalized by ESA. The electrocatalytic methanol oxidation on Pd black and Pd nanosheets were compared by CV measurements in 0.1 M methanol + 0.5 M KOH solution. As shown in Figure S4, the Pd sheets show superior electrocatalytic oxidation to methanol. The maximum current density (derived from peak $a_1$ in Figure S4) is 1.2 mA /cm$^2$ at 0.014 V, which is 150% higher than the Pd black, which shows maximum current density of 0.50 mA/cm$^2$. Note that the peak $a_2$ during the cathodic scan arises from the re-oxidation of the intermediates formed during the anodic scan and/or freshly adsorbed methanol on a reduced PdO surface[5,6].

3. **Free energy calculation for Pd growth between GO sheets.**

The nucleation and growth of Pd between GO sheets can be understood thermodynamically by calculating the total free energy of the Pd, which is the driving force for both nucleation and growth. The free energy of a particle can be defined as the sum of the surface free energy and the bulk free energy. For Pd growth between two GO sheets, the interaction between Pd and GO sheets, as well as the van der Waals adhesion between two GO layers need to be considered. Hence the free energy can be expressed as,

$$\Delta G = Vu + S\gamma + (S_1 + S_2)\gamma' + E_{vdW} \qquad (2)$$

where $V$ is the volume of the Pd nucleus, $u$ is the is the cohesive energy per volume of the nucleus between GO sheets, $S$ is the surface area of the nucleus, $\gamma$ is the surface free energy (surface tension), $S_1$ and $S_2$ are the nucleus' contact areas with the top and bottom GO layers respectively, $\gamma'$ is the interaction parameter (adhesion energy) between GO sheets and Pd, and $E_{vdW}$ is the van der Waals adhesion energy between two GO sheets.

Due to the observed directional growth of Pd between GO sheets, the growth is modelled by assuming that the Pd nucleus is a spherical segment with a radius $r$ and height $z$ as shown in the inset of Figure S5. The crystal free energy of nanoconfined Pd can be defined as $u(z) = u_0 - \alpha\,(z-d)^2$ where $u_0$ is the bulk crystal free energy, $\alpha$ is coefficient, and $d$ is a characteristic length at which Pd interaction with the GO recovers the bulk cohesive energy, i.e. $u(d) = u_0$. Here we assume $d$ = 1-2 nm since the molecular scale interactions become weaker above these limits. The bulk crystal free energy $u_0$ is negative, and depends on the temperature $T$, Boltzmann's constant $k_B$, the supersaturation of the solution $S'$, which is the ratio between the supersaturation concentration $C$ and the electrolyte concentration $C_0$, and the electrolyte molar volume $\Omega$. The $u_0$ is written as:

$$u_0 = -\frac{k_B T}{\Omega}\ln(S') \qquad (3)$$

The coefficient $\alpha$ can be determined as $\alpha \approx \frac{u_0}{d^2}$ by assuming $u(0) \sim 0$ because the nucleation starts at the bottom GO sheets.

Bringing the above definitions together, Eq. (2) can be expressed as

$$\Delta G(r,z) = V u(z) + S\gamma + \pi r^2 [\cos^2(\theta) + 1]\gamma' + E_{vdW}(z), \qquad (4)$$

where, $\pi r^2 (\cos^2[\theta] + 1)$ is the sum of the contacting areas of Pd with the top and bottom graphene sheets, $\theta = \sin^{-1}\left(\frac{z}{r}\right)$, $V = \pi\left(r^2 z - \frac{z^3}{3}\right)$ and $S = 2\pi r z$ are the volume and surface area of the spherical segment as shown in Figure S5 inset.

Figure S5 shows the variation of total free energy of Pd layers as a function of the thickness of Pd calculated using Eq. 4 for different values of $r$. Here, we used a $\Omega \sim$ 187 ml/mol, S $\sim$ 100 [7,8], $d$ = 1.5 nm, $\gamma \sim$ 1.743 J/m$^2$ [9], $\gamma'$ = -0.084 J/m$^2$ [10], and $E_{vdW} = -\frac{H}{12\pi z^2}$ where $H$ is the Hamaker constant, $\sim 0.3\times10^{-19}$J [11]. As shown in Figure S5, irrespective of the value of $r$, the free energy $\Delta G$ is positive above a thickness of ~ 5 nm, indicating the Pd growth between the GO layers is energetically unfavorable.

Here, we used the previously predicted graphene-Pd interaction parameter for our calculation since for GO, an area of 40–60% remains free from functionalization, and hence we assume that $\gamma'$ does not significantly vary from pristine graphene. However, to further understand the importance of the $\gamma'$ on the self-limiting growth of Pd we have calculated the $\Delta G$ for different values of $\gamma'$ by adding a pre-factor to $\gamma'$ and found that to alter the self-limiting growth thickness significantly from 5 nm, the value of $\gamma'$ needs to either increase or decrease by several times. For example, a four time decrease in $\gamma'$ leads to a self-limiting growth thickness of less than 3 nm. To accurately determine the value of $\gamma'$ for different oxygen functional group coverage, further dedicated density functional theory calculations are required.

### 4. ECD of Cu inside the GO laminate

ECD of copper (Cu) inside the GO laminates was performed using the same experimental conditions as used for the ECD of Pd. 1M $CuSO_4$ aqueous solution was used as the electrolyte. Both thin layers of GO on ITO glass and freestanding GO laminates were used for the ECD of Cu. Figure S6a shows the representative AFM image of Cu deposited thin GO flake (3 minutes ECD). Similar to the Pd case, electrochemically deposited Cu also shows two-dimensional flaky structure, however the thicknesses of the individual Cu crystals vary from several to tens of nanometers (Figure S6b and c). To further check the possibility of self-limiting growth of continuous Cu layers between GO sheets, we have also examined a longer time (> 12 hours) ECD of Cu between layers of thick GO laminates (1 μm). Figure S6d shows a SEM image of such a sample after a random mechanical exfoliation. It demonstrates that unlike the case for Pd, Cu islands were overlapping laterally and randomly placed between the GO layers, suggesting the absence of self-limiting growth.

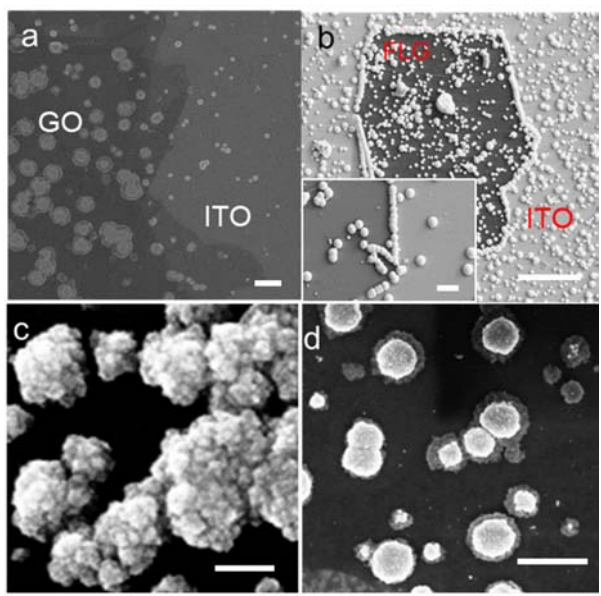

**Figure S1. Electrochemical deposition of Pd on GO, pristine graphene, and rGO covered ITO glass substrates. a-c,** SEM images obtained after the ECD of Pd on a few layer (~ 3 layers) GO covered ITO glass (partially covered), mechanically exfoliated FLG on ITO glass (~ 10 layers), and a few-layer rGO flake (~ 3 layers) on ITO glass with 30 seconds ECD, respectively. Scale bars, **a**. 2 μm, **b**. 10 μm, and **c**. 100 nm respectively. 1b inset: High magnification SEM obtained from the FLG-ITO boundary. Scale bar, 1 μm. **d**, SEM image showing Pd particulate growth out of the GO layers on ITO glass after 3 minutes ECD. Scale bar, 500 nm.

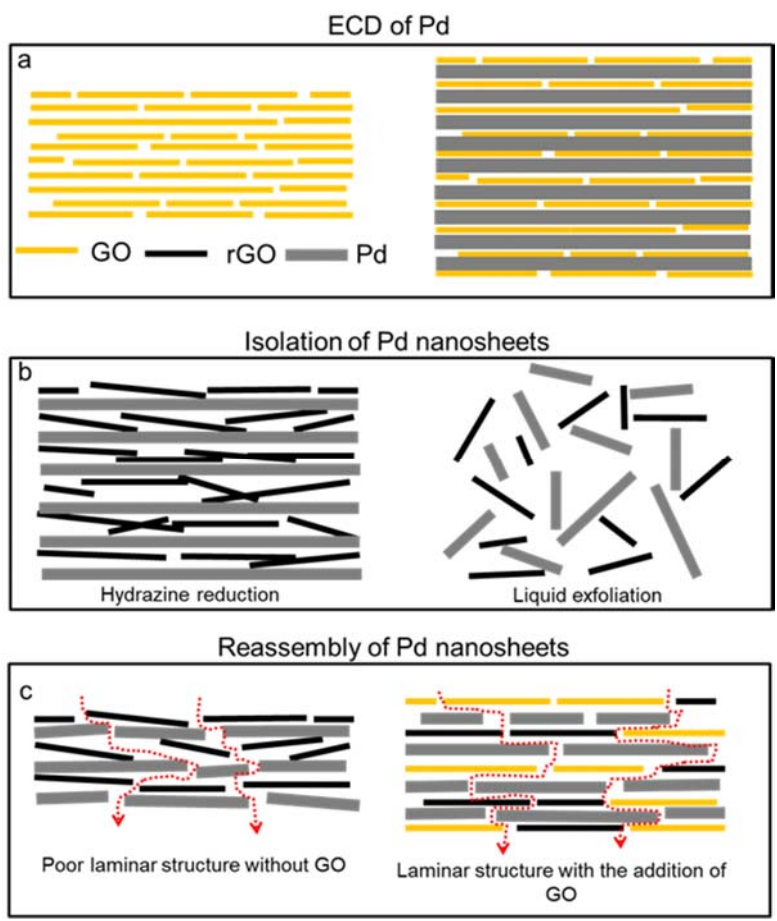

**Figure S2. Schematic showing 2D Pd deposition between GO, their isolation, and reassembly. a**, Schematic of ECD of Pd in the GO laminate, **b**, reduction of Pd-GO by hydrazine, and its liquid exfoliation, and **c**, reassembly of exfoliated Pd nanosheet without and with additional GO sheets. Red dotted lines indicate the molecular permeation pathway.

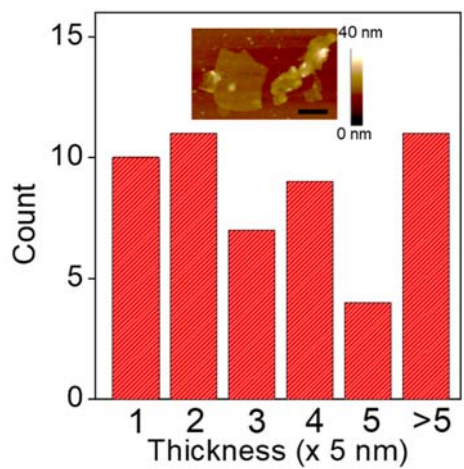

**Figure S3. Characterization of Pd dispersion.** The plot shows the thickness distribution of the exfoliated Pd sheets. Inset: the typical AFM image of the exfoliated Pd nanosheet. Scale bar, 500 nm.

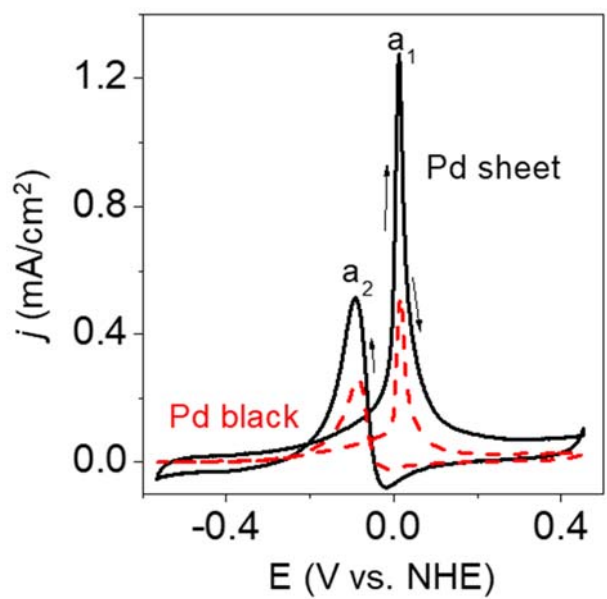

**Figure S4. Electrocatalytic properties of Pd sheets and Pd black for MOR.** The plots are the CV curves recorded in 0.1 M methanol + 0.5 M KOH solution at a scan rate of 50 mV/s. Inset: the CV curve of Pd black electrode in 0.5M $HClO_4$ solution at a scan rate of 50 mV/ s. The grey area highlights the reduction peak of PdO which is used to calculate the ESA.

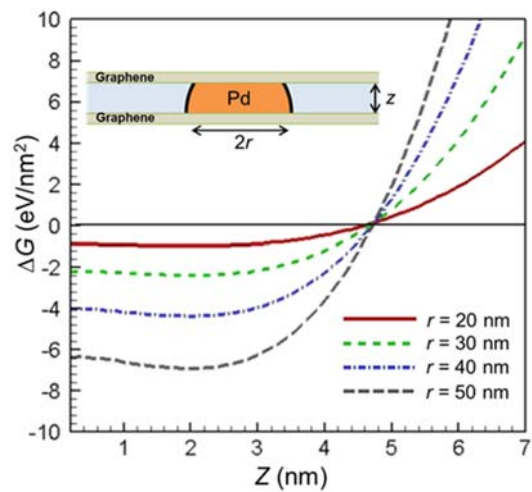

**Figure S5. Free energy for Pd growth between GO sheets.** The total free energy of Pd as a function its thickness ($z$) calculated using Eq. 4 for different values of $r$. Inset: schematic showing a spherical segment of Pd island confined between GO sheets.

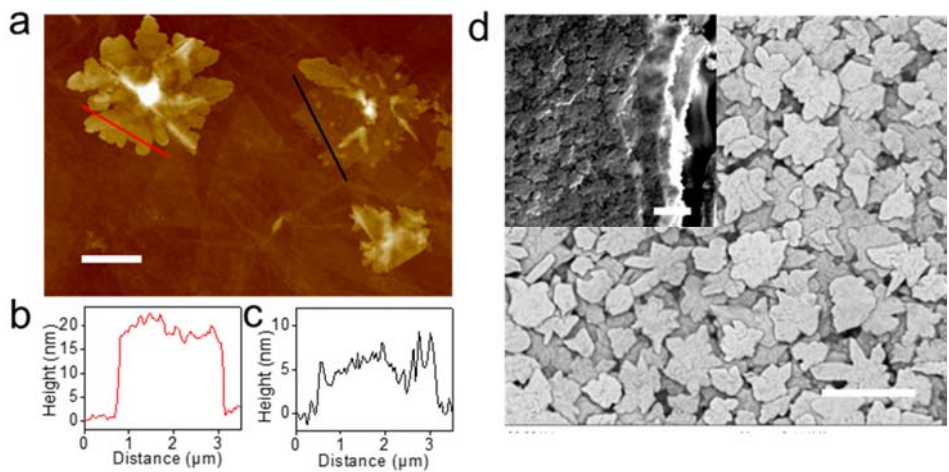

**Figure S6. ECD of Cu inside the GO laminate. a**, AFM image obtained from the Cu deposited few-layer GO on ITO glass. Scale bar, 1μm. **b and c**, the thickness profiles (color coded) obtained along the red and black lines in Figure S6a. **d**, SEM image obtained from a Cu deposited 1 μm thick GO laminate after a random exfoliation. Scale bar, 20 μm. Inset: Low magnification SEM image of the laminate near to the edge of the laminate showing the laminar structure. Scale bar, 20 μm.